\documentclass[aps,pra,reprint,superscriptaddress,longbibliography]{revtex4-2}
\usepackage{amsmath,amssymb,bm,graphicx}
\usepackage[colorlinks=true,linkcolor=blue,citecolor=blue,urlcolor=blue,filecolor=blue]{hyperref}
\hypersetup{
 pdftitle={Collective information backflow from a local population record: Identifiability and learning},
 pdfauthor={Yifan Shi, Pengju Chen, Duan Wang, Quanzhen Ding, Ting Yu}
}
\newcommand{\ket}[1]{\lvert #1\rangle}
\newcommand{\bra}[1]{\langle #1\rvert}

\newcommand{\Diss}{\mathcal D}
\newcommand{\Nback}{\widetilde{\mathcal N}}
\newcommand{\Echan}{\mathcal E}
\newcommand{\eqn}[1]{\hyperref[#1]{Eq.~(\ref*{#1})}}
\newcommand{\tabref}[1]{\hyperref[#1]{Table~\ref*{#1}}}
\newcommand{\figref}[1]{\hyperref[#1]{Fig.~\ref*{#1}}}
\newcommand{\secref}[1]{\hyperref[#1]{Sec.~\ref*{#1}}}
\newcommand{\appref}[1]{\hyperref[#1]{Appendix~\ref*{#1}}}
\begin{document}
\title{Collective information backflow from a local population record:\\ Identifiability and learning}
\author{Yifan Shi}
\author{Pengju Chen}
\author{Duan Wang}
\author{Quanzhen Ding}
\author{Ting Yu}
\email{tyu1@stevens.edu}
\affiliation{Department of Physics and Center for Quantum Science and Engineering, Stevens Institute of Technology, Hoboken, New Jersey 07030, USA}

\begin{abstract}
Information backflow in a composite quantum system is defined through pairs of
evolving states, whereas an experiment may provide only a restricted
population record. We study this mismatch for two identical qubits coupled
collectively to a vacuum reservoir with an exponential correlation function.
An ideal bright-state population trajectory determines the bath-parameter
combinations relevant to a finite-time trace-distance backflow diagnostic
of the full two-qubit channel, evaluated on a fixed time grid and state-pair
bank. The unresolved detuning sign leaves this diagnostic unchanged.
This relation allows sparse population records to predict the diagnostic
using support-vector learning or direct physical fitting. Finite measurement
statistics impose a different restriction: a two-hypothesis bound identifies
reservoirs with similar short population records but substantially different
backflow. For one six-time protocol with 600 total shots, every estimator
using those observations has expected absolute error at least approximately
0.13 at one of two specified reservoirs. Thus a population can be sufficient
within a known bath family without permitting uniformly accurate inference
from a short noisy record.
\end{abstract}
\maketitle

\section{Introduction}

A common reservoir couples the relaxation of different qubits. The resulting
collective decay leaves a dark state unaffected while a bright state can
exchange an excitation with the reservoir. When the reservoir retains memory,
this exchange can restore distinguishability between initially different
system states. The trace-distance measure introduced by Breuer, Laine, and
Piilo quantifies such information backflow by optimizing over the initial
pair~\cite{breuer2009}. For two qubits, a common environment also makes the
relation between different non-Markovianity diagnostics more complicated than
for a single amplitude-damped qubit~\cite{fanchini2013}. Measuring one decay
curve therefore need not directly measure the backflow of the composite
system.

Machine learning offers a way to estimate a dynamical diagnostic from fewer
observations than are needed to evaluate its definition. Fanchini and
collaborators used support-vector regression to estimate non-Markovianity
from a small number of qubit expectation values. They also related the
choice of observation time to parameter identifiability and the measurement
precision needed to distinguish nearby dynamics~\cite{fanchini2021}.
Quantum-kernel methods~\cite{tancara2023} and variational quantum
circuits~\cite{dinani2023} have also been used to estimate
entanglement-based non-Markovianity. Quantum regressors have been
applied to trace-distance backflow in dephasing
dynamics~\cite{huang2026}. The observable and target must therefore
be considered together when interpreting what a sparse record can predict.

The complementary problem is to identify the environment itself. Observable
time traces can constrain parameters of a known Markovian
open-system model~\cite{zhang2015}, and population-based learning has been used in
non-Markovian noise spectroscopy~\cite{scarpetta2025}. Information about an
unknown environment can also be extracted from sequential projective
single-shot measurements by learning a Markovian embedding of the system
dynamics~\cite{luchnikov2020}. For the independently prepared population
records considered here, a particularly relevant comparison is the
reconstruction of Lorentzian spectral-density parameters from
amplitude-damping signals and the finite-time stability analysis of the
forward map~\cite{peleteiro2026}. A bright state in a collectively damped
two-qubit system has the same type of scalar survival dynamics. Inferring
collective backflow, however, requires connecting the information in that
sector to distinguishability evolution under the full two-qubit map.

Restricted readout also raises a statistical question. For non-Markovian
dephasing, likelihood and Bayesian inference can estimate noise parameters,
with measurement times chosen to improve precision~\cite{varona2025}.
Gaussian non-Markovian noise in many-body systems can also be learned using
product-state preparations, local controls, and product observables, with
sample-complexity guarantees~\cite{montana2025}. Such reconstruction
protocols characterize the noise from specified measurements. Here the
quantity to be estimated is a finite-time backflow diagnostic evaluated on
a fixed time grid and state-pair bank, and the available data are a fixed
short population record. Its estimation error need not be captured by
parameter precision alone.

We address this question within a vacuum single-exponential reservoir model.
We show that the ideal bright-state population determines the parameter
combinations needed for the collective diagnostic, which is invariant under
the remaining detuning-sign ambiguity. Scaling and conjugation symmetries
then identify distinct dynamical classes for the learning tests. Six early
noiseless populations support accurate prediction on held-out classes and
additional off-grid parameters. Direct physical fitting also recovers the
noiseless targets and gives smaller mean absolute error in the tested
finite-shot comparisons. Yet two allowed reservoirs can produce similar
early count statistics while having substantially different later backflow.
A target-specific lower bound therefore limits every estimator using the
prescribed six-time record. Together, these results distinguish what a local
population determines within a known model from what a short noisy
measurement can resolve.

\section{Collective relaxation and information backflow}
\label{sec:model}

\subsection{Common exponential reservoir}

We consider two identical qubits with transition frequency $\omega_0$ and
equal coupling to a common vacuum reservoir. In units with $\hbar=1$, the
system Hamiltonian and collective lowering operator are
\begin{equation}
 \begin{aligned}
 H_S&=\omega_0 N,\\
 N&=\sigma_+^{(1)}\sigma_-^{(1)}+\sigma_+^{(2)}\sigma_-^{(2)},\\
 L&=\sigma_-^{(1)}+\sigma_-^{(2)},
 \end{aligned}
 \label{eq:system}
\end{equation}
where $\sigma_-=\ket0\bra1$ and the computational basis is
$(\ket{00},\ket{01},\ket{10},\ket{11})$.
The reservoir is initially factorized from the qubits, with correlation
function
\begin{equation}
 \alpha(t)=\frac{\Gamma\gamma}{2}
 e^{-(\gamma+i\Omega_B)t},\qquad t\geq0.
 \label{eq:correlation}
\end{equation}
Here $\gamma>0$ is the inverse correlation time, $\Omega_B$ is the reservoir
center frequency, and $\Gamma>0$ fixes the coupling scale. The detuning is
$\Delta=\Omega_B-\omega_0$. There is no drive, qubit-frequency mismatch,
thermal excitation, or independent local decay in this model.

A single damped bosonic mode gives a useful exact realization of this
correlation~\cite{garraway1997}. Let $c$ annihilate that mode. The joint state
$R$ obeys
\begin{align}
 \dot R&=-i[H_{\rm pm},R]+2\gamma\Diss[c]R,
 \label{eq:pseudomode}\\
 H_{\rm pm}&=\omega_0N+\Omega_Bc^\dagger c
 +\sqrt{\Gamma\gamma/2}\,(L^\dagger c+Lc^\dagger),\nonumber\\
 \Diss[c]R&=cRc^\dagger-\tfrac12\{c^\dagger c,R\}.
 \nonumber
\end{align}
The mode starts in vacuum. Its free correlation, including its coupling
strength, is precisely \eqn{eq:correlation}; matching such environmental
correlations establishes the corresponding reduced dynamics under the
Gaussian and factorized-state assumptions~\cite{tamascelli2018}.
Physically, the mode stores an excitation before losing it to a memoryless
reservoir. The joint evolution is Markovian, while the reduced two-qubit
evolution need not be.

Excitation exchange and the vacuum initial mode restrict the reachable
joint space to total excitation at most two. Its dimension is eight, even
for an arbitrary initial two-qubit density matrix. We therefore retain the
two-excitation sector rather than approximating the full map by the bright
survival amplitude. The basis and map construction are given in
\appref{app:map}.

\subsection{A sampled trace-distance target}

For a reduced channel $\Echan_t$, the distinguishability of two inputs is
\begin{equation}
 D_{\rho_1,\rho_2}(t)=\tfrac12
 \bigl\|\Echan_t(\rho_1)-\Echan_t(\rho_2)\bigr\|_1.
 \label{eq:distance}
\end{equation}
The Breuer--Laine--Piilo (BLP) measure maximizes its accumulated positive
variation over the initial pair~\cite{breuer2009}. Our numerical target uses
a fixed pair bank $\mathcal B$, a fixed time window, and a finite sampling
grid. With scaled time $\tau=\Gamma t$, it is
\begin{equation}
 \begin{aligned}
 \Nback&=\max_{(\rho_1,\rho_2)\in\mathcal B}
 \sum_{j=0}^{199}\bigl[D_{\rho_1,\rho_2}(\tau_{j+1})\\
 &\qquad-D_{\rho_1,\rho_2}(\tau_j)-\epsilon_D\bigr]_+,
 \qquad \tau_j=0.1j,
 \end{aligned}
 \label{eq:target}
\end{equation}
where $[x]_+=\max(x,0)$ and $\epsilon_D=10^{-7}$ suppresses small numerical
increments. The bank contains four specified orthogonal pure-state pairs
and 50 random orthogonal pairs together with their complex conjugates,
giving 104 pairs. Its construction is specified in \appref{app:bank}.

For exact channel values, this restricted positive-variation sum does not
exceed the continuous-time BLP measure on the same window. A zero value means
that this bank and grid do not detect backflow; it does not establish
divisibility of the channel. Orthogonality of optimal pairs does not justify
replacing the full four-dimensional optimization by pure pairs: mixed states
can be needed in higher dimensions~\cite{wissmann2012}. We consequently learn
the explicitly defined diagnostic \eqn{eq:target}, not an assumed converged
BLP supremum.

\section{What a local population determines}
\label{sec:sufficiency}

\subsection{Dynamical equivalence}

The rotating frame generated by $\omega_0(N+c^\dagger c)$ removes the common
carrier frequency. Define
\begin{equation}
 a=\gamma/\Gamma,\qquad d=\Delta/\Gamma,\qquad \tau=\Gamma t.
 \label{eq:scaled}
\end{equation}
The joint equation becomes
\begin{equation}
 \begin{aligned}
 \frac{dR}{d\tau}&=-i[h_{a,d},R]+2a\Diss[c]R,\\
 h_{a,d}&=d c^\dagger c+\sqrt{a/2}(L^\dagger c+Lc^\dagger).
 \end{aligned}
 \label{eq:scaled-generator}
\end{equation}
Thus the complete rotating-frame channel depends only on $(a,d,\tau)$.
Changing $\Gamma$, $\gamma$, and $\Delta$ by the same factor changes the
physical clock but not this reduced dynamics. A common output rotation does
not change the trace distance.

There is also a conjugation relation. In the real coupling convention of
\eqn{eq:scaled-generator}, complex conjugation changes $h$ to $-h$.
Mode parity $\Pi=(-1)^{c^\dagger c}$ then gives
$\Pi(-h_{a,d})\Pi^\dagger=h_{a,-d}$, while leaving the dissipator and the
initial vacuum invariant. After tracing out the mode,
\begin{equation}
 \Echan^{R}_{a,-d}(\rho^*)=
 \bigl[\Echan^{R}_{a,d}(\rho)\bigr]^*.
 \label{eq:conjugation}
\end{equation}
Because $\mathcal B$ is closed under conjugation, its maximum in
\eqn{eq:target} is unchanged. For the fixed scaled grid,
\begin{equation}
 \Nback=f(a,|d|).
 \label{eq:invariant-target}
\end{equation}
This is an equivalence of the target, not literal equality of all
laboratory-frame Pauli expectation values.

\subsection{Bright preparation and ideal identifiability}

Prepare $\ket{\Psi_+}=(\ket{01}+\ket{10})/\sqrt2$. Since
$L\ket{\Psi_+}=\sqrt2\ket{00}$, its reduced state is
\begin{equation}
 \begin{gathered}
 \rho_+(\tau)=P(\tau)\ket{\Psi_+}\bra{\Psi_+}
                  +[1-P(\tau)]\ket{00}\bra{00},\\
 P=|u|^2.
 \end{gathered}
 \label{eq:bright-state}
\end{equation}
The survival amplitude satisfies
\begin{equation}
 \begin{gathered}
 u''+(a+id)u'+a u=0,\\
 u(0)=1,\qquad u'(0)=0,
 \end{gathered}
 \label{eq:bright-ode}
\end{equation}
where primes denote scaled-time derivatives. Choosing
$Z=\ket0\bra0-\ket1\bra1$, one local measurement gives
\begin{equation}
 P(\tau)=1-\langle I\otimes Z\rangle_\tau.
 \label{eq:readout}
\end{equation}
The protocol requires an entangled bright-state preparation, but neither
two-qubit state tomography nor a joint measurement at readout.

An ideal population record near $\tau=0$ identifies the combinations in
\eqn{eq:invariant-target}. Differentiating \eqn{eq:bright-ode} with respect
to the scaled time $\tau=\Gamma t$, at fixed $a$ and $d$, gives
\begin{align}
 \left.\frac{d^2P}{d\tau^2}\right|_{\tau=0}&=-2a,\nonumber\\
 \left.\frac{d^3P}{d\tau^3}\right|_{\tau=0}&=2a^2,\nonumber\\
 \left.\frac{d^4P}{d\tau^4}\right|_{\tau=0}
 &=8a^2-2a^3+2ad^2.
 \label{eq:derivatives}
\end{align}
Hence, for finite $a>0$,
\begin{equation}
 \begin{aligned}
 a&=-\frac12\left.\frac{d^2P}{d\tau^2}\right|_{\tau=0},\\
 d^2&=\frac{1}{2a}\left.\frac{d^4P}{d\tau^4}\right|_{\tau=0}
      -4a+a^2.
 \end{aligned}
 \label{eq:identification}
\end{equation}
The second and fourth population derivatives at $\tau=0$ determine $a$
and $|d|$, and therefore determine $\Nback$; the third derivative is fixed
by $a$. This is a structural identifiability statement within the known
reservoir family. It does not prescribe differentiation of noisy data or
prove that any six discrete values can be inverted uniquely.

The missing detuning sign is physically meaningful. The whole state in
\eqn{eq:bright-state} is identical for $d$ and $-d$, whereas a preparation
coherent between $\ket{00}$ and $\ket{\Psi_+}$ retains a phase that changes
under conjugation. The population therefore identifies an invariant target
more readily than it identifies the channel. Equally, a different reservoir
family can contain information invisible to this preparation. The inference
uses the model in \secref{sec:model}, rather than a model-independent
population criterion for non-Markovianity.

The numerical observation times below are specified in $\tau$ and therefore
assume a calibrated $\Gamma$. Ideal physical-time derivatives can separate
$\Gamma$, $\gamma$, and $|\Delta|$ as well; this distinct identification
problem is described in \appref{app:derivatives}.

\section{Prediction from short population records}
\label{sec:prediction}

\subsection{Observation sets and distinct dynamics}

The ideal trajectory identifies the target, but a finite record need not
contain enough information for accurate prediction. We therefore restrict
the input to six early population values,
\begin{equation}
 \begin{aligned}
 \bm x_6={}&\bigl(P(0.25),P(0.5),P(1),\\
 &\qquad P(1.5),P(2),P(3)\bigr).
 \end{aligned}
 \label{eq:six-times}
\end{equation}
It ends well before the target window $\tau=20$. We compare it with twelve
nonzero-time populations extending to $\tau=20$, and with 780 scalar Pauli
features from four probe states at thirteen times. The latter contains all
15 nonidentity two-qubit Pauli expectations per probe and time. It is an
information-rich feature baseline, not a count of independent experimental
settings: compatible observables can share measurements, and initial values
are already fixed by preparation. A fourth baseline uses the known reduced
bath parameters, which are additional information rather than a measurement
protocol for an unknown bath.

We start from 1134 parameter triples with $\omega_0=1$. Their ranges are
$0.05\leq\Gamma\leq0.6$, $0.1\leq\gamma\leq5$, and
$0\leq\Omega_B\leq2$. The actual grid is specified in
\appref{app:learning}. Applying \eqn{eq:invariant-target} leaves 447 distinct
$(a,|d|)$ classes, of which 166 have positive sampled backflow. We retain one
representative per class, choosing nonnegative detuning in the original grid
order without using its label.

Grouping only by the sign of the detuning would still allow rescaled copies
of the same reduced dynamics to occur in both training and testing. We split
the $(a,|d|)$ classes instead, so that the test set probes dynamics absent from
training and validation.

\subsection{Learning and off-grid prediction}

A support-vector classifier distinguishes zero from positive labels, and a
support-vector regressor learns $\log_{10}\Nback$ on positive training
examples. Both use radial-basis-function kernels. Standardization is fitted
on training examples only; regularization and the classification threshold
are selected on validation examples using the rule in \appref{app:learning}.
The deployed estimate is zero when the
classifier predicts zero, and is the exponentiated regression output
otherwise.

\tabref{tab:clean} compares the observation sets. Six population values give
accurate predictions under the same learning procedure that is applied to
the larger input vectors. Additional scalar features do not guarantee better
finite-sample regression: they also change the distance geometry seen by the
kernel. The table therefore supports sparse prediction in this model, not an
information-theoretic superiority of six observations over twelve or 780.
The positive-label logarithmic score tests the regressor separately; the
all-label mean absolute error also includes classification mistakes.

\begin{table*}[tbp]
\centering
\caption{Prediction on distinct dynamical classes with noiseless inputs.
Entries are means and sample standard deviations across ten splits; these
spreads describe split sensitivity, not population confidence intervals.
$R^2_+$ is evaluated for the regressor on truly positive test labels in
logarithmic units. MAE is the mean absolute error of the deployed two-stage
estimate on all test labels.}
\label{tab:clean}
\begin{ruledtabular}
\begin{tabular}{lccc}
Inputs & $F_1$ & $R^2_+$ & MAE\\
\hline
Six populations & $0.948\pm0.026$ & $0.959\pm0.026$ & $0.0053\pm0.0028$\\
Twelve populations & $0.947\pm0.022$ & $0.946\pm0.047$ & $0.0050\pm0.0035$\\
780 Pauli features & $0.940\pm0.049$ & $0.895\pm0.062$ & $0.0075\pm0.0023$\\
Two known bath parameters & $0.963\pm0.046$ & $0.946\pm0.029$ & $0.0031\pm0.0016$\\
\end{tabular}
\end{ruledtabular}
\end{table*}

\begin{figure}[tbp]
\centering
\includegraphics[width=\columnwidth]{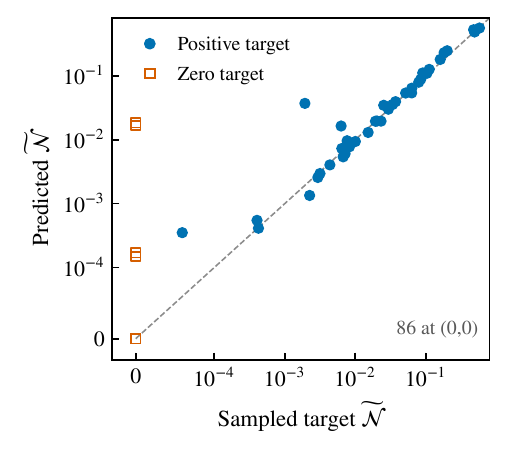}
\caption{Off-grid prediction from six noiseless early populations. All 128
baths are shown: blue circles denote positive targets and orange open squares
denote zero targets. Of the latter, 86 points coincide at the origin and four
have positive predictions. The target is defined by \eqn{eq:target}; the
dashed line marks equality with the prediction. Both axes are linear at
values up to $10^{-4}$ and logarithmic above, retaining exact zeros without
shifting or flooring the data.}
\label{fig:offgrid}
\end{figure}

To test points outside the original grid, we generate 128 additional
parameters in the convex hull of the original
$(\log_{10}a,\log_{10}(1+|d|))$ coordinates. The selection is made without
their target values, and the seed-101 training and validation choices remain
fixed. The six-population estimator gives $F_1\simeq0.95$, positive-label
logarithmic $R^2\simeq0.92$, and all-label mean absolute error $0.0028$.
The corresponding error of a zero predictor is $0.027$.
The pointwise comparison in \figref{fig:offgrid} includes the zero-target
baths and their four false-positive predictions. Thus the successful
prediction is not confined to held-out entries of the original parameter
grid. This is interpolation in the stated reduced-coordinate domain; it is
not extrapolation to another correlation function.

\subsection{Direct physical fitting}

The identifiability result also supplies a non-learning alternative. We fit
the analytic solution of \eqn{eq:bright-ode} directly to the same six
populations, then evaluate \eqn{eq:target} for the fitted parameters. The
fit uses only the observed values and the parameter domain, not training
labels. A fixed multistart search treats noiseless observations by least
squares and finite-shot observations by binomial maximum likelihood.

On the 68 seed-101 test classes, the noiseless physical fit reproduces the
target to numerical accuracy. This result matters
for interpreting the learning experiment: when the model is known, learning
is not required to make the noiseless inference. It provides an amortized
map from observations to the target, while physical fitting explicitly
reconstructs and propagates a member of the model family. Neither procedure
has access to information absent from the observations.

\section{Finite-shot information limits}
\label{sec:noise}

\subsection{The local readout experiment}

For the state \eqn{eq:bright-state}, the probability that the measured qubit
is excited is $q_j=P(\tau_j)/2$. With $n$ independent shots at each of the
six times in \eqn{eq:six-times}, the observations are
\begin{equation}
 K_j\sim\operatorname{Binomial}(n,q_j),\qquad
 \widehat P_j=2K_j/n.
 \label{eq:shots}
\end{equation}
The total budget is $6n$ preparations and local measurements. Shots at
different times are taken on independently prepared systems, rather than
through repeated interventions on one reservoir trajectory.

For the matched comparison in \tabref{tab:noise}, the physical fit and the
learned predictor receive exactly the same noisy test records. The learned
predictor is trained and validated with independently generated records at
the same shot budget. We then keep each predictor fixed and generate 50
independent test-record realizations on the same 68 classes. Physical fitting
has a smaller mean all-label absolute error at every displayed budget;
paired uncertainty estimates are given in Appendix~\ref{app:learning}.
This ordering is metric-dependent:
logarithmic errors emphasize small positive labels and need not rank the
estimators in the same way. No universal advantage of the learning procedure
is implied by the noiseless sparse-feature comparison.

\begin{table*}[tbp]
\centering
\caption{Mean absolute error on the same 68 seed-101 test classes, using
the same six-time readout records. The learning column includes both stages
of the predictor. Finite-shot entries are mean $\pm$ sample standard deviation
over 50 paired test-record realizations, conditional on the fixed test classes
and trained predictors. Each realization uses the displayed shots per time
and class; records from different realizations are not pooled.}
\label{tab:noise}
\begin{ruledtabular}
\begin{tabular}{rcc}
Shots per time & Physical fit & Learned predictor\\
\hline
Noiseless & $<10^{-9}$ & $0.0024$\\
100 & $0.01568\pm0.00386$ & $0.02520\pm0.00625$\\
1000 & $0.00470\pm0.00114$ & $0.01108\pm0.00189$\\
10000 & $0.00178\pm0.00053$ & $0.00749\pm0.00133$\\
\end{tabular}
\end{ruledtabular}
\end{table*}

\subsection{An estimator-independent target-error bound}

Errors in a particular regressor do not by themselves identify a measurement
limit. The relevant obstruction is physical: two environments can yield
nearly indistinguishable observations while producing different amounts of
later backflow. A record compatible with both cannot support an accurate
target estimate for both. To quantify this obstruction, consider two
allowed bath parameters $\theta_0=(a_0,|d_0|)$ and
$\theta_1=(a_1,|d_1|)$, with target separation
$\delta=|f(\theta_0)-f(\theta_1)|$. Let $p_i(\bm K)$ be the joint count
distribution under $\theta_i$. Their statistical affinity is
\begin{equation}
 \begin{aligned}
 A_n&=\sum_{\bm K}\sqrt{p_0(\bm K)p_1(\bm K)}\\
 &=\prod_{j=1}^{6}\Bigl[\sqrt{q_{0j}q_{1j}}\\
 &\qquad+\sqrt{(1-q_{0j})(1-q_{1j})}\Bigr]^n.
 \end{aligned}
 \label{eq:affinity}
\end{equation}
The binomial theorem gives the second equality exactly. Summarizing the
shots by their six counts loses no information about these parameters.

For any estimate $\widehat f(\bm K)$, define its expected absolute error
at hypothesis $i$ by $R_i=\mathbb E_i|\widehat f-f(\theta_i)|$.
Integrating the triangle inequality against the smaller of the two
probabilities gives
\begin{align}
 R_0+R_1
 &\geq\delta\sum_{\bm K}\min\{p_0(\bm K),p_1(\bm K)\}\nonumber\\
 &=\delta\,[1-\operatorname{TV}(p_0,p_1)],
 \label{eq:overlap-risk}
\end{align}
where $\operatorname{TV}(p_0,p_1)=\frac12\sum_{\bm K}|p_0-p_1|$ is the
total variation distance. The affinity inequality
$\operatorname{TV}\leq\sqrt{1-A_n^2}$~\cite{fuchs1999} therefore yields
\begin{equation}
 \max\{R_0,R_1\}\geq
 \frac{\delta}{2}\left(1-\sqrt{1-A_n^2}\right).
 \label{eq:risk-bound}
\end{equation}
The proof also permits randomized estimators and an independently generated
training set: adding the same independent law under both hypotheses does
not improve their distinguishability. It does not permit extra measurements
or parameter-revealing side information.

\begin{figure*}[tbp]
\centering
\includegraphics[width=\textwidth]{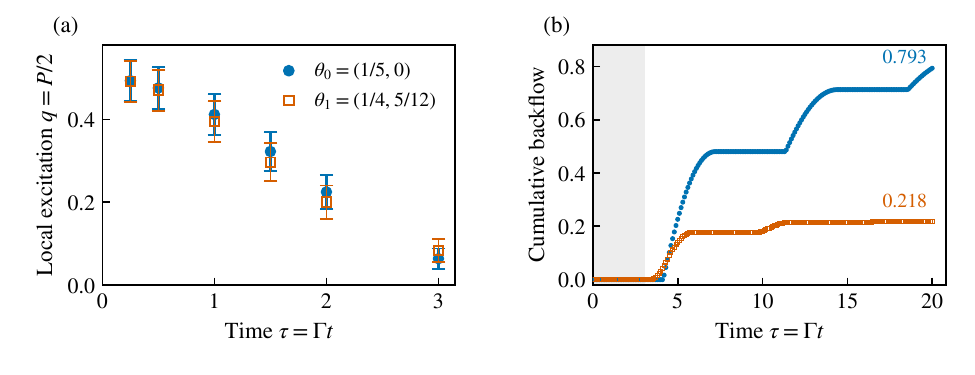}
\caption{Early readout and later return for $\theta_0=(1/5,0)$ (blue circles)
and $\theta_1=(1/4,5/12)$ (orange open squares). (a) Expected local excitation
fractions $q_j=P(\tau_j)/2$ at the six observation times; bars show
$\pm\sqrt{q_j(1-q_j)/100}$, the theoretical standard deviation of $K_j/100$
for 100 shots per time. (b) Accumulated positive trace-distance increments
for the fixed pair $\ket{01}$ versus $\ket{10}$, using the $0.1$ grid and
threshold of \eqn{eq:target}. This pair maximizes the final-horizon sum for
both baths. Markers show the discrete accumulation; the gray band marks
the readout window $0\leq\tau\leq3$. The endpoints $0.793$ and $0.218$ give
the respective full-bank sampled targets.}
\label{fig:hardpair}
\end{figure*}

An informative pair in the present parameter set is
\begin{equation}
 \theta_0=(1/5,0),\qquad \theta_1=(1/4,5/12).
 \label{eq:hard-pair}
\end{equation}
\figref{fig:hardpair} compares their early observation statistics with the
later accumulated return. The two sampled targets are approximately $0.793$
and $0.218$, although the six early populations have similar binomial
statistics. At $n=100$, $A_n\simeq0.843$, and \eqn{eq:risk-bound} gives
$\max\{R_0,R_1\}\gtrsim0.133$ using the numerically evaluated target
separation. Both hypotheses may be realized with the same calibrated
$\Gamma$, so their physical observation times need not reveal which bath
was used. Thus, with 600 total shots at the prescribed six times, every
estimator has expected absolute error at least approximately $0.133$ for
at least one of these two environments. The affinity quantifies overlap
of the complete count distributions; the conclusion does not rely only
on the visual proximity of the mean populations or their error bars.

The example establishes that the measurement record itself can limit
prediction, even when the reservoir family and an exact forward model are
known. It does not rank learning and physical fitting: both are subject to
the same bound. The risk here is the larger expected error at two fixed
baths, whereas \tabref{tab:noise} averages errors over a test set. The pair
was selected from the existing parameter classes to exhibit this
limitation, not to prove that it is globally least distinguishable.
Measuring later or changing the shot allocation can improve the
distinguishability of the environments, but changes the observation
protocol to which the bound must be applied.

These conclusions use two qubits coupled to one common vacuum reservoir.
The common coupling already correlates the qubits' relaxation. The ideal identification formula relies on the more specific
assumptions of equal couplings and a single-exponential correlation.
At finite temperature, absorption and thermal occupation would have to be
included, and the bright-state survival equation and exact restriction to
at most two total excitations could no longer be used as written.
For more general temporal or spatial bath correlations, one must likewise
establish which channel parameters are identifiable from the chosen
population record. Independently prepared binary measurements can still
have binomial statistics, but their probabilities and the backflow target
must be obtained from the extended dynamics. The two-hypothesis inequality
then applies to those observation laws and target values; the present
numerical bound does not carry over without that calculation.

\section{Discussion and conclusion}

The population in \eqn{eq:bright-state} describes a single excitation,
whereas the target in \eqn{eq:target} tests the complete two-qubit channel.
The connection between them is a model-identification statement: the
population fixes the parameter combinations of that channel which affect
the target. It is not an equality between population revival and the sampled
backflow in \eqn{eq:target}. For example, the bank contains the pair
$\ket{01},\ket{10}$, whose trace distance is $|u|=\sqrt P$. Its sampled
positive variation is a directly accessible lower diagnostic. At
$(a,|d|)=(1/3,10/3)$ it gives approximately $0.046$, whereas the full-bank
value is $0.059$: this single contrast misses about $22\%$ of the sampled
backflow. Other initial pairs reveal distinguishability return not captured
by that contrast. Model knowledge links their evolution to the measured
population without making the observables identical.

The same distinction explains why detuning-sign ambiguity need not prevent
prediction. The population does not resolve that sign, but the target is
constant on the conjugate pair of dynamics. An estimator should therefore
learn a function of physically distinguishable classes relevant to its
target, rather than of every parameter label used to generate data. This
also changes what counts as a held-out test. A change in dimensional rates
need not produce an unseen reduced trajectory when the observation clock is
scaled with those rates.

Sparse backflow inference already connects observation time with
identifiability and measurement precision~\cite{fanchini2021}, while
population-based spectroscopy examines stability and parameter
conditioning~\cite{peleteiro2026}. In the collective problem, the additional
connection is between the identifiable population parameters and a diagnostic
of the complete map. Parameter recovery is therefore already sufficient for
noiseless prediction, making physical fitting an essential comparison.
Learning provides a direct surrogate from observations to the target; its
value depends on finite-data behavior, repeated-inference cost, and available
prior information. The present comparison supports accurate sparse
prediction, but no general accuracy advantage over physical fitting.

Finite shots put a separate condition on this conclusion. The ideal
identification formula uses exact derivatives; in a short-time expansion,
the detuning first enters the population at fourth order. Distinct baths
can thus remain statistically close over an early observation window even
when their later information return differs strongly. Non-Markovian
tomography quantifies parameter precision~\cite{varona2025}, while
many-body noise-learning protocols give sample-complexity guarantees under
Gaussian assumptions~\cite{montana2025}.
The bound in \eqn{eq:risk-bound} instead concerns expected error in the
collective backflow diagnostic itself, for every estimator using the
specified six counts. It applies at finite shot number without converting a
parameter error into a target error or assuming an efficient estimator.
Improving a regressor cannot remove this measurement limitation.

For larger qubit registers, the next question is which preparations and
local observables identify the parameters relevant to collective backflow.
Additional collective sectors, unequal couplings, and more general temporal
or spatial noise correlations may require more than one population record.
Within an identifiable model, observation times and shot allocations can
then be chosen to distinguish environments with different target values.
Extending this design to thermal and non-exponential reservoirs would test
how the required readout changes with the environmental assumptions.

For collective relaxation in an exponential reservoir, an ideal local
population record thus contains enough information to determine the chosen
backflow diagnostic. Sparse records support useful prediction on distinct
dynamics, but finite-shot records can leave baths with substantially different
backflow difficult to distinguish. Measurement times, shot budget, and prior
model knowledge together determine the inference available to either a
learned surrogate or a physical fit.

\appendix
\section{Complete-map representation and numerical checks}
\label{app:map}

The joint basis for \eqn{eq:scaled-generator} is
\begin{equation}
 \begin{aligned}
 (&\ket{00,0},\ket{00,1},\ket{00,2},\ket{01,0},\\
  &\ket{01,1},\ket{10,0},\ket{10,1},\ket{11,0}),
 \end{aligned}
 \label{eq:joint-basis}
\end{equation}
where the last index is the mode occupation. Exchange preserves total
excitation, and the Lindblad jump can only lower it. This is therefore an
exact invariant reachable space, not a low-excitation approximation to an
initial state containing higher sectors.

Column-vectorizing the joint density matrix gives a constant $64\times64$
generator. We embed the sixteen system matrix units with a vacuum mode,
propagate all columns together, and take the mode partial trace to obtain
the $16\times16$ system map. For the uniform target grid, one matrix
exponential gives the step propagator; repeated multiplication generates
all sampled times. This recurrence introduces no time-discretization
approximation to the constant generator. Discretization remains in the
definition of the backflow target.

For comparison, we also use the stochastic-propagator construction of
Ding~\cite[Chap.~2, Eqs.~(2.5), (2.23), and (2.24)]{ding2021stochastic},
which obtains the reduced map by Gaussian
averaging products of noise-expansion coefficients. In the present vacuum
two-qubit model, excitation conservation terminates that expansion at noise
order two. The exponential-kernel recurrence then closes on fourteen
$16\times16$ auxiliary superoperators $\mathcal L_{m,n,k}$, with
$0\leq m,n\leq2$ and $0\leq k\leq\min(m,n)$. Here $m$ and $n$ are the
ket- and bra-side noise orders, and $k$ counts their completed Gaussian
contractions. Initially only $\mathcal L_{0,0,0}=I_{16}$ is nonzero; in
column-vectorized form the reduced map is
$\Echan_t=\mathcal L_{0,0,0}+\mathcal L_{1,1,1}
+\mathcal L_{2,2,2}/2$. This finite closure follows from the vacuum and
excitation-conservation assumptions, rather than a numerical cutoff.

With the kernel convention in \eqn{eq:correlation}, the physical-time
decay term is
$-[(m-k)(\gamma+i\Omega_B)+(n-k)(\gamma-i\Omega_B)]\mathcal L_{m,n,k}$.
Both noise orders therefore enter with a minus sign. We do not use the
positive bra-side signs printed in Eqs.~(2.38) and (2.39) of
Ref.~\cite{ding2021stochastic}.

We compare this hierarchy with the independently assembled damped-mode map
at $(\Gamma,\gamma,\Omega_B)=(0.5,0.5,1)$, $(0.1,5,0)$,
$(0.05,0.1,2)$, and $(0.2,0.4,1)$, in units with $\omega_0=1$.
At each point the comparison times are
$\tau\in\{0,0.25,1,4,20\}$. The hierarchy is integrated with DOP853,
using relative and absolute tolerances $10^{-10}$ and $10^{-12}$;
the damped-mode map is evaluated by direct matrix exponentiation.
The maximum absolute matrix-element discrepancy is $2.1\times10^{-7}$.
More directly relevant to learning, recomputing all 447 representative
labels with the damped-mode map gives maximum absolute label discrepancy
$2.2\times10^{-9}$, and no disagreement between zero and positive labels.
These checks justify treating the numerical label error as small relative
to the reported learning prediction errors. They do not estimate the missing
continuous-time or unrestricted-pair contribution to the BLP measure.

\section{Pair bank and full-observation baseline}
\label{app:bank}

The four deterministic pairs are $(\ket{00},\ket{11})$,
$(\ket{01},\ket{10})$, $(\ket{\Phi_+},\ket{\Phi_-})$, and
$(\ket{\Psi_+},\ket{\Psi_-})$, where
$\ket{\Phi_\pm}=(\ket{00}\pm\ket{11})/\sqrt2$ and
$\ket{\Psi_\pm}=(\ket{01}\pm\ket{10})/\sqrt2$.
For each additional pair, a complex Gaussian $4\times4$ matrix is
QR-decomposed, its diagonal phases are corrected, and the first two columns
give orthogonal Haar-distributed pure states. We use 50 such pairs with
random seed 12345 and include their 50 complex-conjugate pairs.

The 780-feature baseline uses the probes $\ket{\Phi_+}$,
$\ket{\Psi_+}$, $\ket{11}$, and
$\frac34\ket{\Psi_-}\bra{\Psi_-}+\frac1{16}I_4$.
For each probe it retains all nonidentity products of
$I,\sigma_x,\sigma_y,Z$ at times
\begin{equation*}
 \begin{aligned}
 \tau={}&0,0.25,0.5,1,1.5,2,3,\\
 &4,6,8,12,16,20.
 \end{aligned}
\end{equation*}
The twelve-population baseline uses the same times except zero. The
six-population vector is fixed by \eqn{eq:six-times}; it was not chosen by
optimizing test scores.

\section{Identifiability and short-time sensitivity}
\label{app:derivatives}

Writing $s=a+id$, the amplitude derivatives at zero follow recursively from
\eqn{eq:bright-ode}:
\begin{equation*}
 \begin{gathered}
 u''(0)=-a,\qquad u'''(0)=as,\\
 u^{(4)}(0)=a^2-as^2.
 \end{gathered}
\end{equation*}
Using $P=u^*u$ gives \eqn{eq:derivatives}, including the contribution
$6|u''|^2$ to its fourth derivative. Equivalently,
\begin{equation}
 \begin{aligned}
 P(\tau)={}&1-a\tau^2+\frac{a^2}{3}\tau^3\\
 &+\frac{4a^2-a^3+ad^2}{12}\tau^4+O(\tau^5).
 \end{aligned}
 \label{eq:population-expansion}
\end{equation}
This expansion is local at finite $a,d$. The numerical six-time schedule is
not uniformly in this short-time regime over the full parameter range.

For a single local Bernoulli shot with probability $q=P/2$, the Fisher
information for signed detuning is
$I_{dd}=(\partial_d q)^2/[q(1-q)]$.
At early time and near resonance,
\begin{equation}
 I_{dd}=\frac{a^2d^2\tau^8}{36}+\text{higher-order terms}.
 \label{eq:fisher-local}
\end{equation}
The sign ambiguity is reflected in the vanishing sensitivity at $d=0$.
Using $\vartheta=d^2$ retains the leading fourth-order sensitivity, although
$\vartheta=0$ remains a boundary of the parameter domain $\vartheta\geq0$.
For $n_j$ shots at times $\tau_j$, let $S_k=\sum_jn_j\tau_j^k$. Eliminating an unknown
nuisance parameter $a$ from the leading Fisher matrix yields
\begin{equation}
 I_{\vartheta\vartheta}^{\rm eff}\simeq
 \frac{a^2}{144}\left(S_8-\frac{S_6^2}{S_4}\right).
 \label{eq:fisher-nuisance}
\end{equation}
Distinct times are needed to separate the leading responses. These local
parameter sensitivities explain a conditioning problem, but do not by
themselves establish a rate for estimating $\Nback$. The main-text
two-hypothesis bound instead addresses that target directly at the actual
sampling times.

For physical-time derivatives, write
$P_t^{(k)}(0)\equiv\left.d^kP(t)/dt^k\right|_{t=0}$ and set
$K=\Gamma\gamma$. Then $P_t''(0)=-2K$, $P_t'''(0)=2K\gamma$, and
$P_t^{(4)}(0)=8K^2-2K\gamma^2+2K\Delta^2$. Thus
\begin{equation*}
 \begin{aligned}
 K&=-\frac{P_t''(0)}2,\\
 \gamma&=-\frac{P_t'''(0)}{P_t''(0)},\\
 \Gamma&=K/\gamma,\\
 \Delta^2&=\frac{P_t^{(4)}(0)}{2K}-4K+\gamma^2.
 \end{aligned}
\end{equation*}
This ideal continuous-record reconstruction does not remove the clock
calibration used to implement the chosen scaled sampling schedule.

\section{Learning, fitting, and off-grid sampling}
\label{app:learning}

The dimensional source grid is the Cartesian product
\begin{align*}
 \Gamma&\in\{0.05,0.075,0.1,0.15,0.2,\\
 &\qquad 0.3,0.4,0.5,0.6\},\\
 \gamma&\in\{0.1,0.15,0.2,0.25,0.3,0.35,\\
 &\qquad 0.4,0.5,0.75,1,1.5,2,3,5\},\\
 \Omega_B&\in\{0,0.25,0.5,0.75,1,\\
 &\qquad 1.25,1.5,1.75,2\}.
\end{align*}
Equivalence classes are formed from exact ratios of these decimal grid
values, rather than by rounding the labels. Each class split is stratified
by positive versus zero labels. The test fraction is $0.15$; a second split
of the remaining pool gives a $0.15$ validation fraction of the original
class set, giving 312 training, 67 validation, and 68 test classes.
Seeds 101--110 are used for the outer split and the following
integer for each inner split.

The classifier regularization is selected from $C\in\{1,10,100\}$.
For each candidate, thresholds are zero and 19 validation-score quantiles
with probabilities $0.05,0.10,\ldots,0.95$. We choose the threshold that
maximizes validation $F_1$. The regressor uses
$C\in\{1,10,100\}$ and insensitive-tube width
$\varepsilon\in\{0.03,0.1\}$, maximizing validation $R^2$ for positive
logarithmic labels. Kernel widths follow the standardized training-data
variance. The known-parameter baseline uses
$(\log_{10}a,\log_{10}(1+|d|))$. No test example determines a scaler,
hyperparameter, or classification threshold.

The physical fit searches $1/6\leq a\leq100$ and $0\leq|d|\leq20$ in
coordinates $(\log a,\operatorname{arsinh}|d|)$. Eight separated starts are
chosen from a fixed, label-free $33\times81$ dictionary by observation
residual. Bounded least squares minimizes either the population residual or
the binomial deviance. Each start has at most 350 residual evaluations,
with function, step, and gradient tolerances $10^{-12}$. This multistart
procedure is a numerical comparator, not a proof of global optimization.
All fitted parameters are propagated through the same full-map target.
For the 68 noiseless test records, the maximum absolute target discrepancy
is below $10^{-9}$.

For the finite-shot comparison in \tabref{tab:noise}, the budget-specific
predictors are fixed after training and validation on the original noisy
records. At each budget, 50 independent realizations of the 68 test records
are generated and used identically by both estimators. We first average
absolute errors over the 68 classes within each realization, then summarize
these 50 paired means. The table reports their sample standard deviations,
not standard errors. The Monte Carlo standard error of each mean is its
sample standard deviation divided by $\sqrt{50}$; it does not include
retraining, split variability, or uncertainty over a population of environments.

The mean learner-minus-fit MAE differences are $0.00952$, $0.00638$, and
$0.00570$ for $n=100$, $1000$, and $10000$, respectively. Their approximate
pointwise 95\% Student-$t$ intervals, with 49 degrees of freedom, are
$[0.00797,0.01106]$, $[0.00577,0.00698]$, and $[0.00529,0.00612]$.
These intervals are not simultaneous across budgets. All 10200 paired
records are retained. Of the 81600 optimizer starts, 36 reached the
evaluation limit; none was selected as the lowest-deviance fit.

For the off-grid test, a Latin hypercube with seed 6201 supplies 256 points
in the bounding rectangle of the logarithmic reduced coordinates. Of
these, 218 lie in the source convex hull; the first 128 in the generated
order are retained before computing their labels. None coincides with a
source point. One lies outside the image of the original dimensional
parameter box, illustrating that its image need not equal a convex hull
after a nonlinear coordinate transformation. All are valid points of the
same exponential-reservoir model. Training and selection remain fixed to
the seed-101 partition.

\bibliography{references}
\end{document}